\documentclass[%
reprint,
amsmath,amssymb,
aps,
pra,
twocolumn,
floatfix,
]{revtex4-2}

\usepackage{float}
\usepackage{comment}
\usepackage{physics}
\usepackage{graphicx}
\usepackage{dcolumn}
\usepackage{bm}

\begin{document}
    
    \title{Topological protection in the collisional gates using qubit doublons}
    
    \author{Guanjie Xu}
    \thanks{These authors contributed equally.}
    \affiliation{State Key Laboratory of Surface Physics and Department of Physics, Fudan University, Shanghai 200433, China}
    
    \author{Chufan Zhang}
    \thanks{These authors contributed equally.}
    \affiliation{State Key Laboratory of Surface Physics and Department of Physics, Fudan University, Shanghai 200433, China}
    
    \author{Xiaoyu Zeng}
    \affiliation{State Key Laboratory of Surface Physics and Department of Physics, Fudan University, Shanghai 200433, China}
    
    \author{Zijie Zhu}
    \email{zhuzijie@fudan.edu.cn}
    \affiliation{State Key Laboratory of Surface Physics and Department of Physics, Fudan University, Shanghai 200433, China}
    
    \date{\today}

	\begin{abstract}
		Geometric quantum gates offer a route to robust quantum control, but geometry alone does not guarantee protection against control errors. Here we establish the topological origin of the geometric phase in the qubit-doublon collisional gates demonstrated by Kiefer et al. [Nature 652, 609 (2026)], showing that spatial inversion symmetry alone is sufficient to protect this geometric phase at arbitrary interaction strength. We further clarify the working regime of entangling gates based on direct exchange interaction and demonstrate that they remain functional even at strong interactions. Finally, we develop a unified framework that compares the robustness of three major collisional-gate protocols: adiabatic exchange, nonadiabatic exchange, and superexchange. Our results connect topological protection with the practical requirements of quantum information processing, providing a basis for designing robust collisional gates in neutral-atom quantum processors.

	\end{abstract}
	
	\maketitle
	
	
	\section{\label{sec:level1}Introduction}

    Controlled collisions between neutral atoms support two-qubit operations~\cite{Jaksch1999,Calarco2000,Mandel2003,Anderlini2007,Kaufman2015}, including entangling gates in the \(\mathrm{SWAP}^{\alpha}\) family~\cite{LossDiVincenzo1998,Hayes2007}. Recent experiments have demonstrated high-fidelity collisional gates in optical lattices~\cite{Yang2020,Bojovic2026,Kiefer2026}. Extending this performance across large arrays requires robustness against lattice inhomogeneity and fluctuations in tunnelling and interaction strength.
    Geometric and holonomic gates encode quantum operations in the geometry of an evolution path~\cite{Berry1984,WilczekZee1984,Zanardi1999,Sjoqvist2012}. They can reduce sensitivity to certain control errors, with the robustness depending on the chosen path and the error model~\cite{ZhuZanardi2005,Xu2017RobustPaths}. For a fixed path in state space, the geometric phase is independent of the traversal rate~\cite{AharonovAnandan1987,Sjoqvist2015}, but geometric evolution alone does not guarantee improved robustness against control errors~\cite{Colmenar2022}. Symmetry-protected topological quantization provides a stronger constraint by fixing the phase through a discrete invariant~\cite{HughesProdanBernevig2010}.
    
    In the demonstrated qubit-doublon exchange gate~\cite{Kiefer2026}, the geometric phase implements a $\mathrm{SWAP}$ operation while interactions generate a tunable dynamical phase that enables entanglement. At zero interaction, the quantization of the geometric phase was associated with time-reversal and chiral symmetries and with adiabatic following of an exact dark state. Yet, three questions are still open: whether the geometric phase remains quantized at finite interactions; whether the gate can be extended beyond the experimentally demonstrated weak-interaction regime for faster entangling operations; and how its robustness compares with nonadiabatic exchange~\cite{Yang2020}, extending the reported comparison with superexchange. Recent proposals for fast collisional gates further motivate this comparison~\cite{Jensen2019,NemirovskySagi2021,Singh2025,Weill2026,Reuter2026}.
    
    In this work, we show that spatial inversion symmetry alone is sufficient to fix the geometric phase to $\pi$ even when finite interactions break chiral symmetry and remove the dark state. This phase encodes a \(\mathbb{Z}_2\) topological invariant that remains nontrivial at any finite interaction strength. An auxiliary three-band stub-lattice representation identifies the geometric phase with the quantized Zak phase, connecting the gate to symmetry-protected band topology~\cite{Zak1989,HughesProdanBernevig2010}. Beyond phase quantization, we clarify the working regime of direct-exchange entangling gates, showing that sufficiently wide bias sweeps allow operation even at strong interactions, where the narrowing spectral gap increases the adiabatic timescale. We also develop a unified framework for comparing the robustness of adiabatic exchange, nonadiabatic exchange, and superexchange under static interaction and tunnelling errors. Adiabatic exchange is the most robust of the three protocols against static tunnelling errors, suggesting an advantage since tunnelling is particularly sensitive to lattice depth fluctuations~\cite{Bloch2008} and is an important source of error in current lattice implementations~\cite{Kiefer2026}. 
	
	\section{Double-well Hamiltonian and its spatial symmetry}
	
	We consider a system of two spin-$1/2$ fermionic atoms with opposite spins trapped within a double-well potential formed by an optical superlattice, as sketched in Fig.~1(a). 
    Assuming that only the lowest orbital of each site is populated, the relevant two-particle Hilbert space is spanned by the single-occupancy configurations $\ket{\uparrow,\downarrow}$ and $\ket{\downarrow,\uparrow}$, together with the doublon states $\ket{\uparrow\downarrow,0}$ and $\ket{0,\uparrow\downarrow}$.
	The system is governed by three experimentally tunable parameters: the tunnelling amplitude $t$, which couples configurations that differ by the transfer of one atom between the wells; the on-site interaction $U$, which sets the energy cost of double occupation; and the left-right bias $\Delta$, which determines the energy offset between the two wells~\cite{Murmann2015,Chalopin2025}.
	
	The two-particle Hilbert space can be decomposed into sectors of definite total spin. For the exchange dynamics, we introduce the  spin-singlet and spin-triplet states
	\begin{equation*}
		\ket{s}
		=
		\frac{\ket{\uparrow,\downarrow}
			-
			\ket{\downarrow,\uparrow}}{\sqrt{2}},
		\qquad
		\ket{t_0}
		=
		\frac{\ket{\uparrow,\downarrow}
			+
			\ket{\downarrow,\uparrow}}{\sqrt{2}},
	\end{equation*}
	together with the symmetric and antisymmetric doublon states
	\begin{equation*}
		\ket{D_\pm}
		=
		\frac{\ket{\uparrow\downarrow,0}
			\pm
			\ket{0,\uparrow\downarrow}}{\sqrt{2}}.
	\end{equation*}
	In the ordered basis
	$\{\ket{D_+},\ket{D_-},\ket{s},\ket{t_0}\}$, the Hamiltonian reads
	\begin{equation}
		H(\Delta)
		=
		\begin{pmatrix}
			U & 2\Delta & -2t & 0\\
			2\Delta & U & 0 & 0\\
			-2t & 0 & 0 & 0\\
			0 & 0 & 0 & 0
		\end{pmatrix}.
		\label{eq:effective_exchange_hamiltonian}
	\end{equation}
    The triplet state $\ket{t_0}$ is decoupled from the singlet sector spanned by $\{\ket{D_+},\ket{D_-},\ket{s}\}$, where nontrivial dynamics occurs. Nevertheless, we retain $\ket{t_0}$ explicitly because it provides a phase reference for the evolution of the singlet component. 
    For qubits encoded in the spin states of individual atoms, a relative phase accumulated by $\ket{s}$ with respect to the triplet states, $\ket{s}\rightarrow e^{i\pi\alpha}\ket{s}$, directly implements a partial two-qubit swap gate $(\mathrm{SWAP})^\alpha$.
    The doublon states are not populated at the beginning and the end of the gate operation, ensuring no leakage outside of the qubit computational space.
	Within the singlet sector, the bias $\Delta$ drives the transitions between the even and odd doublon states $\ket{D_\pm}$, while the tunnelling $t$ couples $\ket{s}$ and $\ket{D_+}$, with the interaction energy $U$ setting the detuning [Fig.~1(b)]. This interpretation follows from identifying the relevant singlet block of the Hamiltonian as the effective Hamiltonian of a driven three-level system under the rotating-wave approximation\cite{Vitanov2017}.
	
	\begin{figure}[!t]
		\centering
		\includegraphics[width=0.48\textwidth]{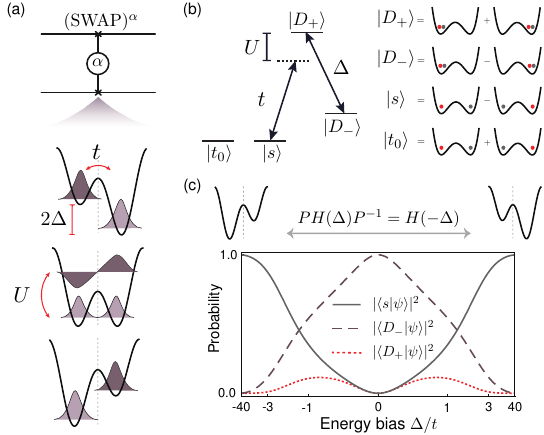}
		\caption{Physical setting and symmetry structure of the two-qubit gates realized by populating qubit doublon states.
        (a) A $(\mathrm{SWAP})^\alpha$ gate is realized in a dynamical double-well model, where two spin-$1/2$ fermionic qubits occupy two distinct spatial orbitals (gray and brown). The double-well potential is characterized by the tunnelling amplitude $t$, the bias $\Delta$, and the on-site interaction $U$. The bias is swept from a large negative to a large positive value, leading to wavefunction overlap of the two qubits during the gate operation.
        (b) The double-well Hamiltonian, expressed in the basis $\{\ket{D_+},\ket{D_-},\ket{s},\ket{t_0}\}$, is illustrated in a $\Lambda$-system-like representation, where the triplet $\ket{t_0}$ is decoupled from the other three basis states. 
        (c) The Hamiltonian exhibits a spatial inversion symmetry, $P H(\Delta) P^{-1} = H(-\Delta)$, where $P$ is the inversion operator that exchanges the two wells and sends $\Delta \rightarrow -\Delta$. The lower panel plots the state composition of the relevant eigenstate (namely $\ket{s}$ at the beginning and the end) during the gate operation under finite interactions. In contrast to the $U=0$ case, it is no longer a dark state and acquires an admixture of the bright state $\ket{D_+}$.}
		\label{fig:double_well_model}
	\end{figure}
	
	The Hamiltonian has spatial inversion symmetry, which exchanges the left and right wells and reverses the energy bias [Fig.~\ref{fig:double_well_model}(c), upper panel]. 
	In the basis $\{\ket{D_+},\ket{D_-},\ket{s},\ket{t}\}$, the parity operator is represented by $P=\text{diag}(1,-1,1,-1)$, and the Hamiltonian obeys
	\begin{equation}
		P H(\Delta)P^{-1}
		=
		H(-\Delta).
		\label{eq:spatial_inversion_relation}
	\end{equation}
    This symmetry is sufficient to protect the quantization of the geometric phase, as we discuss below. We emphasize that it is preserved at any Hubbard $U$, which ensures that the protection holds at any interaction strength.

    \section{Quantization of the geometric phase}

	The two-qubit $(\mathrm{SWAP})^\alpha$ gate is realized by adiabatically sweeping the bias $\Delta$ from a large negative value to a large positive value. We introduce an angular parameter $\theta$ through
	\begin{equation}
		\cot\frac{\theta}{2}
		=
		-\frac{\Delta}{t},
		\qquad
		\theta\in(0,2\pi),
		\label{eq:theta_parameter}
	\end{equation}
    which map the spatial inversion $\Delta\rightarrow-\Delta$ to
	$\theta\rightarrow2\pi-\theta$, and Equation~\eqref{eq:spatial_inversion_relation}
	can be written as
	\begin{equation}
		PH(\theta)P^{-1}
		=
		H(2\pi-\theta).
		\label{eq:spatial_inversion_theta}
	\end{equation}
	Under this parametrization, the $\Delta$ sweep corresponds to $\theta$ from $0$ to $2\pi$.
    Here, we choose a periodic gauge in which the relevant eigenstate $\ket{\psi(\theta)}$ followed during the gate operation approaches the singlet $\ket{s}$ in both limits, thereby forming a closed path in the state space. 
    The geometric phase accumulated when adiabatically following this relevant eigenstate is given by~\cite{Berry1984,AharonovAnandan1987}
	\begin{equation}
		\gamma_g = \int_0^{2\pi} i\braket{\psi(\theta)}{\partial_{\theta}\psi(\theta)}d\theta.
		\label{eq:geometric_phase}
	\end{equation}
	This inversion symmetry relates the geometric phase to the parity eigenvalues $\xi(\theta)$ at the two inversion-invariant points $\theta=0,\pi$ through
    \begin{equation}
    e^{i\gamma_g}=\xi(0)\xi(\pi)=(-1)^\nu, 
    \label{eq:geometric_phase_invariant}
    \end{equation}
    where $\nu\in\{0,1\}$ is an inversion-protected $\mathbb{Z}_2$ topological invariant associated with the closed eigenstate path.
    The parity product is the single-branch form of the standard inversion invariant~\cite{HughesProdanBernevig2010}.
    Its value remains unchanged as long as the inversion symmetry is preserved.
    A derivation of Eq.~\eqref{eq:geometric_phase_invariant} adapted to our system is provided in Appendix~\ref{app:geometric_phase_inversion}.
    For the followed eigenstate $\ket{\psi(\theta)}$, these eigenvalues
    are $\xi(0)=+1$ and $\xi(\pi)=-1$, corresponding
    to $\ket{s}$ and $\ket{D_-}$, respectively.
    Consequently,
    \begin{equation}
    e^{i\gamma_g}=-1 
    \quad \Rightarrow \quad
    \gamma_g=\pi\pmod{2\pi}.
    \label{eq:quantized_geometric_phase}
    \end{equation}

    Crucially, this quantization remains valid for finite Hubbard interaction \(U\). As illustrated in Fig.~\ref{fig:double_well_model}(c), a finite interaction removes the exact dark state, and the instantaneous eigenstate acquires a nonzero admixture of the bright state \(\ket{D_+}\) during the bias sweep. Nevertheless, \(\ket{D_-}\) remains an exact odd-parity eigenstate at \(\Delta=0\) ($\theta=\pi$), independent of \(U\), while the followed eigenstate approaches the even-parity state \(\ket{s}\) in the large-bias limit. The parity eigenvalues, and hence both the geometric phase and the topological invariant, remain unchanged. This establishes that the topological protection of the geometric phase does not rely on the existence of an exact dark state.
    In Appendix~\ref{app:real_structure}, we give an alternative proof of this quantization using the real structure of the Hamiltonian.
    Appendix~\ref{app:geometric_phase_numerics} provides numerical confirmation that the geometric phase remains $\gamma_g=\pi$ at finite interaction strengths.

    \section{Stub-lattice representation}

\begin{figure}[!t]
    \centering
    \includegraphics[width=0.48\textwidth]{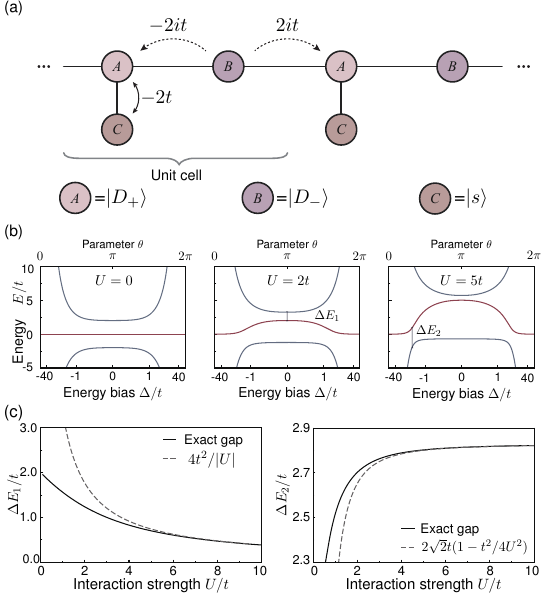}
    \caption{(a) Stub-lattice representation of the double-well Hamiltonian.
    Each unit cell contains three sites. (b) Instantaneous eigenenergies of the singlet-sector Hamiltonian as functions of the dimensionless bias $\Delta/t$ for $U=0$, $2t$, and $5t$; finite tunnelling opens a central gap $\Delta E_1$ and outer gaps $\Delta E_2$ along the branch followed during the sweep (red line). The upper horizontal axis shows the angular parameter $\theta$, identified with the quasimomentum $k$ of the stub lattice. These spectra can therefore also be interpreted as the band spectra of the lattice model. (c) Dependence of these two gaps on the dimensionless interaction strength $U/t$, comparing the exact numerical gaps with the large-$U$ estimates $\Delta E_1\simeq4t^2/U$ and $\Delta E_2\simeq2\sqrt{2}\,t(1-t^2/4U^2)$. The central gap shrinks at large $U/t$.}
    \label{fig:lattice_schematic}
    \end{figure}

    The quantized geometric phase derived above can also be understood as the Zak phase of a one-dimensional lattice. To establish this connection, we identify the angular parameter $\theta$ with a synthetic quasimomentum $k\equiv\theta$. In the ordered singlet-sector basis $\{\ket{D_+},\ket{D_-},\ket{s}\}$, the Hamiltonian takes the Bloch-like form
    \begin{equation}
        H_{\mathcal S}(k)
        =
        \begin{pmatrix}
        U & -2t\cot(k/2) & -2t\\
        -2t\cot(k/2) & U & 0\\
        -2t & 0 & 0
        \end{pmatrix}.
        \label{eq:singlet_bloch_hamiltonian}
    \end{equation}
    A Fourier representation of $H_{\mathcal S}(k)$ corresponds to a one-dimensional stub lattice, as illustrated in Fig.~\ref{fig:lattice_schematic}(a). Each unit cell contains three sites, $A$, $B$, and $C$, corresponding to $\ket{D_+}$, $\ket{D_-}$, and $\ket{s}$, respectively. The momentum-dependent coupling between $\ket{D_+}$ and $\ket{D_-}$ generates long-range imaginary hoppings between the $A$ and $B$ sites, whereas the coupling between $\ket{D_+}$ and $\ket{s}$ gives rise to a real intracell hopping of amplitude $-2t$. The Hubbard interaction $U$ enters as an on-site energy shift of sites $A$ and $B$. Figure~\ref{fig:lattice_schematic}(b) shows the instantaneous eigenenergies of $H_{\mathcal S}(k)$ during the bias sweep for different values of $U$. These spectra can also be interpreted as the three bands of the stub lattice. The relevant eigenstate followed during the exchange gate corresponds to the middle band, which is flat at $U=0$ and becomes dispersive for finite $U$. The detailed mapping is presented in Appendix~\ref{app:lattice_mapping}.

    Under the identification $k\equiv\theta$, the closed exchange path maps onto the synthetic Brillouin zone, with $\ket{\psi(\theta)}$ corresponding to the cell-periodic Bloch eigenstate $\ket{u(k)}$ of the middle band. The geometric phase in Eq.~\eqref{eq:geometric_phase} therefore coincides with the Zak phase of this band~\cite{Zak1989},
    \begin{equation}
        \gamma_g
        =
        \int_{\rm BZ}
        i\braket{u(k)}{\partial_k u(k)}\,dk
        =
        \gamma_{\rm Zak}.
        \label{eq:geometric_zak_equivalence}
    \end{equation}
    As in the SSH model~\cite{SuSchriefferHeeger1979}, inversion symmetry protects the quantization of the Zak phase of the stub lattice~\cite{HughesProdanBernevig2010}. The middle band thus exhibits nontrivial $\mathbb{Z}_2$ topology, which persists at finite $U$ despite the breaking of chiral symmetry and the emergence of band dispersion.

    \section{Extending direct-exchange gates beyond the weak-interaction regime}

    In the weak-interaction regime ($U \le t$), the first-order energy shift induced by $U$ provides the tunable dynamical phase required for direct-exchange entangling gates~\cite{Kiefer2026}. This mechanism readily extends to strong interactions because the middle eigenstate remains gapped from the rest of the spectrum for any finite $U$ [Fig.~\ref{fig:lattice_schematic}(b)], which permits adiabatic tracking of the state. Consequently, the protocol requires only that the singlet component map to this middle branch at the sweep boundaries, rather than that $U$ remain a small perturbation to $t$.
    
    For $U>0$, Fig.~\ref{fig:lattice_schematic}(b) shows that the middle eigenstate asymptotically approaches the bare singlet $\ket{s}$ away from the avoided crossing with the lowest branch. This avoided crossing occurs near $\Delta = -U/2$, where the unperturbed singlet energy intersects the doublon energy $U + 2\Delta$. Tunnelling hybridizes these states and opens the gap $\Delta E_2$. To ensure the initial state cleanly maps onto the middle eigenstate, the starting bias must lie well outside this hybridization region. Symmetrically, the sweep must terminate well past the corresponding crossing near $\Delta = U/2$. Generalizing to either sign of $U$, this requires
    \begin{equation}
    |\Delta_{\mathrm{end}}| - |U|/2 \gg t
    \label{eq:gate_endpoint_condition}
    \end{equation}
    at both sweep boundaries. Consequently, any finite interaction can be accommodated simply by choosing a sufficiently wide bias range.

    With this boundary condition satisfied, the sweep speed is constrained by the need to track the middle branch adiabatically~\cite{Jansen2007}. For large $U/t$, the outer gap $\Delta E_2$ saturates at $2\sqrt{2}t$, whereas the central gap $\Delta E_1 \simeq 4t^2/\vert{}U\vert{}$ shrinks and sets the adiabatic bottleneck [Fig.~\ref{fig:lattice_schematic}(c)].
    This gap also governs the pair-tunnelling gates based on superexchange in symmetric double wells~\cite{Bojovic2026}. 
	
	\section{Comparison of gate robustness}

\begin{figure}[!t]
		\centering
		\includegraphics[width=0.48\textwidth]{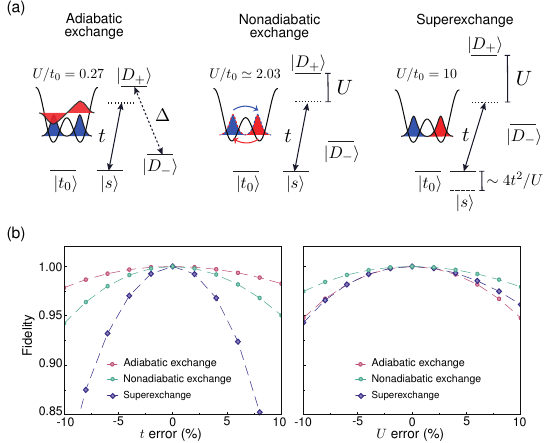}
		\caption{Comparison of collisional $\sqrt{\mathrm{SWAP}}$ gate protocols.
        (a) Schematic level diagrams for adiabatic exchange driven by a bias sweep $\Delta$ (left), nonadiabatic exchange through coherent $\ket{s}$--$\ket{D_+}$ oscillations (middle), and superexchange mediated by virtual doublon occupation with exchange energy $J_{\mathrm{ex}}\simeq4t^2/U$ (right).
        (b) Fidelity measure $F_8^{1/8}$ versus relative static errors in $t$ (left) and $U$ (right), expressed in percent, where $F_8$ is the state fidelity after eight gate repetitions relative to the ideal sequence.
        All protocols use Blackman tunnelling pulses with a common peak amplitude $t_0$, with the working points listed in Table~\ref{tab:gate_working_points}.
        The pulse durations are held fixed when the errors are applied.}
		\label{fig:gate_robustness}
	\end{figure}

    To assess the robustness of the direct-exchange gate discussed above, we compare this adiabatic exchange protocol with nonadiabatic exchange and superexchange protocols implementing the same entangling $\sqrt{\mathrm{SWAP}}$ gate. Using the same Hamiltonian, Eq.~\eqref{eq:effective_exchange_hamiltonian}, we can analyse and compare the three collisional gate mechanisms shown in Fig.~\ref{fig:gate_robustness}(a) within a common coherent-control framework. The adiabatic exchange protocol is a cyclic analogue of stimulated Raman adiabatic passage (STIRAP)~\cite{Vitanov2017}. Nonadiabatic exchange operates in a symmetric double well, $\Delta=0$, where $\ket{D_-}$ decouples and the states $\ket{s}$ and $\ket{D_+}$ form a coupled two-level system with $U$ setting the detuning. The singlet undergoes coherent Rabi oscillations between these states and returns with a phase relative to the triplets that can be tuned through $U$. Superexchange uses the same coupling in the far-detuned regime, $U\gg t$, where irtual occupation of the doublon state lowers the singlet energy by $4t^2/U$~\cite{Trotzky2008}, analogous to an AC Stark shift. During the tunnelling pulse, the instantaneous eigenstate remains approximately $\ket{s}$, accumulating the corresponding dynamical phase.

    For the comparison, we use the same Blackman tunnelling envelope with a common peak amplitude $t_{0}$, while choosing the interaction strength and pulse duration separately for each mechanism (Appendix~\ref{app:robustness}). 
    For the adiabatic exchange protocol, strong interactions can enable rapid dynamical-phase accumulation, but the speed remains limited by the adiabatic requirement. Weak interactions therefore provide a favourable regime for reaching the target phase while remaining adiabatic. We set $U/t_{\max}=0.27$, matching the experimental value in Ref.~\cite{Kiefer2026}, and adjust the range of the linear bias ramp and the pulse duration to achieve the target phase.
    For nonadiabatic exchange, population return and the target phase of $\pi/2$ must be satisfied simultaneously. For a square tunnelling pulse, this occurs at a series of magic ratios, beginning with $U/t=4/\sqrt{3}\approx2.31$~\cite{Yang2020}. We note although the square-pulse gate is purely geometric, population return requires the tunnelling to be switched off precisely at a revival, making it sensitive to pulse timing. 
    A smooth Blackman envelope reduces this sensitivity and yields a more robust implementation (Fig.~\ref{fig:first nonadiabatic}). The chosen working point $U/t_{0}\simeq2.03$ is still close to the first magic ratio.  Additional working points lie near higher magic ratios, but as shown in Fig.~\ref{fig:two nonadiabatic}, they offer lower robustness.
    For superexchange, we choose $U/t_{\max}=10$ to operate within the far-detuned regime; increasing $U$ further reduces the exchange rate and lengthens the gate.

    We assess robustness against static errors in either $U$ or the tunnelling amplitude. For the input $\ket{\uparrow,\downarrow}$, the ideal output is
    \(
    \ket{\Psi_{\mathrm{id}}}
    =\frac{e^{i\pi/4}}{\sqrt{2}}
    \left(\ket{\uparrow,\downarrow}-i\ket{\downarrow,\uparrow}\right).
    \) 
    We define the state fidelity as
    \begin{equation}
    F=\left|\braket{\Psi_{\mathrm{id}}}{\Psi_{\mathrm{out}}}\right|^2,
    \label{eq:fidelity_defintaion}
    \end{equation}
    where $\ket{\Psi_{\mathrm{out}}}$ is the full evolved state. This definition accounts for both relative-phase errors and residual doublon population (see Appendix~\ref{app:robustness}).
    Fig.~\ref{fig:gate_robustness}(b) shows that adiabatic exchange is the most robust against tunnelling errors, whereas nonadiabatic exchange is the most robust against interaction errors at the selected working points. The adiabatic exchange gate retains the highest fidelity over the displayed range of tunnelling fluctuations. Superexchange is particularly sensitive because its phase-accumulation rate depends quadratically on $t$. For nonadiabatic exchange, tunnelling errors perturb both the coherent return condition and the acquired phase. Nonadiabatic exchange also offers a substantially shorter gate duration (Table~\ref{tab:gate_working_points}). However, the practical comparison must account for the different magnitudes of the experimental control errors. In deep optical potentials, tunnelling depends exponentially on the square root of the trap depth, whereas the on-site interaction varies only algebraically~\cite{Bloch2008}. Fluctuations in the lattice depth therefore produce much larger relative changes in $t$ than in $U$. For experiments dominated by these fluctuations, the stronger robustness of adiabatic exchange against tunnelling errors favours high-fidelity operation despite its longer duration.

	\section{Conclusion and outlook}
	
	We have shown that an inversion-protected topological invariant fixes the geometric phase at any finite interaction strength. Direct-exchange entangling gates remain viable at strong interactions, although the narrowing adiabatic gap favours weak interactions in practice. Our unified comparison of adiabatic exchange, nonadiabatic exchange, and superexchange guides protocol selection according to the dominant experimental errors. Nonadiabatic exchange offers faster operation, whereas adiabatic exchange is favoured when tunnelling fluctuations dominate.
    
    Optical lattices provide scalable trapping for large atom arrays with reduced optical-power requirements~\cite{Tao2024,Norcia2024}, but independently addressed entangling gates remain challenging. Combining lattice confinement with local tweezer manipulation~\cite{Gyger2024,Young2024} offers a route to locally controlled collisional gates~\cite{Weitenberg2011Architecture}. The exchange gate studied here is suited to this architecture because its robustness against tunnelling variations could ease the requirements on tweezer-intensity stability and uniformity. A sufficiently broad bias sweep can also accommodate static bias offsets caused by positioning errors, potentially relaxing the alignment requirements between tweezers and lattice sites.
    
    Beyond individual gates, Thouless pumping provides a route to coherent atom transport and entanglement distribution~\cite{Zhu2025,HuKeLee2020}. Combining this transport with collisional gates has enabled circuits for digital programming of spin correlations in fermionic optical lattices~\cite{Kiefer2026SpinCorrelations}. Together with local single-qubit control~\cite{Weitenberg2011Addressing,Barnes2022}, these capabilities could support programmable quantum circuits across large atom arrays.
	
	\begin{acknowledgments}
    We thank Tilman Esslinger and Konrad Viebahn for helpful discussions and comments on the manuscript. This work was supported by Fudan University and the Xianghui Institute.
	\end{acknowledgments}
	
	\appendix
	\setcounter{figure}{0}
    \renewcommand{\thefigure}{A\arabic{figure}}
	
	\section{Proof of $\mathbb{Z}_2$ geometric phase based on inversion symmetry}
	\label{app:geometric_phase_inversion}
	In this appendix, we give a detailed derivation of the quantized geometric phase for the adiabatic exchange gate. Since the triplet sectors is decoupled and has no dynamics, we restrict the derivation to the Hamiltonian in the singlet sector:
	\begin{equation}
		H_{\mathcal S}(\Delta)
		=
		\begin{pmatrix}
			U & 2\Delta & -2t\\
			2\Delta & U & 0\\
			-2t & 0 & 0
		\end{pmatrix},
		\qquad
		\{\ket{D_+},\ket{D_-},\ket{s}\}.
		\label{eq:appC_H_delta}
	\end{equation}
	The bias parameter is compactified as
	\begin{equation}
		\cot\frac{\theta}{2}
		=
		-\frac{\Delta}{t},
		\qquad
		\theta\in(0,2\pi).
		\label{eq:appC_compactification}
	\end{equation}
	The Hamiltonian diverges at the endpoints of this interval. For fixed $t>0$ and finite $U$, the spectral projector $P_{\rm ad}(\theta)=\ket{u_{\rm ad}(\theta)}
	\bra{u_{\rm ad}(\theta)}$ onto the middle eigenstate satisfies
    \begin{equation}
    \lim_{\theta\to0^+}P_{\rm ad}(\theta)
    =
    \lim_{\theta\to2\pi^-}P_{\rm ad}(\theta)
    =
    \ket{s}\bra{s}.
    \label{eq:projector_limits}
    \end{equation}
    Identifying the endpoints therefore defines a closed loop of eigen-projectors, with $P_{\rm ad}(0)=P_{\rm ad}(2\pi)=\ket{s}\bra{s}$.
	The left-right inversion of the double well is represented in the basis
	$\{\ket{D_+},\ket{D_-},\ket{s}\}$ by
	\begin{equation}
		P
		=
		\mathrm{diag}(1,-1,1).
		\label{eq:appC_P}
	\end{equation}
	Since the bias changes sign under left-right exchange, the Hamiltonian satisfies
	\begin{equation}
		P H_{\mathcal S}(\theta) P^{-1}
		=
		H_{\mathcal S}(2\pi-\theta).
		\label{eq:appC_inversion_theta}
	\end{equation}
    
	Let $\ket{u(\theta)}$ be the normalized eigenstate of the nondegenerate branch
	followed in the main text,
	\begin{equation}
		H_{\mathcal S}(\theta)\ket{u(\theta)}
		=
		E(\theta)\ket{u(\theta)}.
		\label{eq:appC_eigen_theta}
	\end{equation}
    We choose a smooth complex gauge on the compactified loop, with
    \begin{equation}
        \ket{u(0)}=\ket{u(2\pi)}=\ket{s}.
    \label{eq:gauge_choice}
    \end{equation}
    The eigenvectors are allowed to be complex even though the Hamiltonian is real.
	Because of Eq.~\eqref{eq:appC_inversion_theta}, the state $P\ket{u(\theta)}$ is an
	eigenstate of $H_{\mathcal S}(2\pi-\theta)$ with the same energy. If this branch is nondegenerate,
	then $P\ket{u(\theta)}$ must be proportional to $\ket{u(2\pi-\theta)}$. Therefore,
	there exists a real phase function $\beta(\theta)$ such that
	\begin{equation}
		P\ket{u(\theta)}
		=
		e^{i\beta(\theta)}
		\ket{u(2\pi-\theta)}.
		\label{eq:appC_sewing_theta}
	\end{equation}
	For the integration below, we choose a continuous real branch of $\beta(\theta)$.
	
	The Berry connection along the compactified loop is defined as
	\begin{equation}
		A(\theta)
		=
		i\braket{u(\theta)}{\partial_\theta u(\theta)}.
		\label{eq:appC_Berry_connection_theta}
	\end{equation}
	We now derive its transformation under inversion. Let
	\begin{equation}
		\theta' = 2\pi-\theta.
	\end{equation}
	Since
	\begin{equation}
		\partial_{\theta'}
		=
		-\partial_\theta,
	\end{equation}
	we have
	\begin{align}
		A(2\pi-\theta)
		&=
		i\braket{u(2\pi-\theta)}{\partial_{\theta'}u(\theta')}
		\nonumber\\
		&=
		-i\braket{u(2\pi-\theta)}{\partial_{\theta}u(2\pi-\theta)}.
		\label{eq:appC_A_transform_start}
	\end{align}
	Using Eq.~\eqref{eq:appC_sewing_theta},
	\begin{align}
		\partial_\theta\ket{u(2\pi-\theta)}
		&=
		\partial_\theta
		\left[
		e^{-i\beta(\theta)}P\ket{u(\theta)}
		\right]
		\nonumber\\
		&=
		-i(\partial_\theta\beta)
		e^{-i\beta(\theta)}
		P\ket{u(\theta)}\notag\\
		&\quad
		+e^{-i\beta(\theta)}
		P\partial_\theta\ket{u(\theta)}.
		\label{eq:appC_derivative_sewing}
	\end{align}
	Substituting Eq.~\eqref{eq:appC_derivative_sewing} into
	Eq.~\eqref{eq:appC_A_transform_start}, and using $P^\dagger P=1$, gives
	\begin{align}
		A(2\pi-\theta)
		&=
		-i
		\left[
		-i\partial_\theta\beta(\theta)
		+
		\braket{u(\theta)}{\partial_\theta u(\theta)}
		\right]
		\nonumber\\
		&=
		-\partial_\theta\beta(\theta)
		-
		i\braket{u(\theta)}{\partial_\theta u(\theta)}
		\nonumber\\
		&=
		-A(\theta)-\partial_\theta\beta(\theta).
		\label{eq:appC_A_transform_theta}
	\end{align}
	In this periodic gauge, the geometric phase is therefore
	\begin{align}
		\gamma
		&=
		\int_0^{2\pi} A(\theta)\,d\theta
		\nonumber\\
		&=
		\int_0^\pi
		\left[
		A(\theta)+A(2\pi-\theta)
		\right]d\theta
		\nonumber\\
		&=
		-\int_0^\pi
		\partial_\theta\beta(\theta)\,d\theta
		\nonumber\\
		&=
		\beta(0)-\beta(\pi)
		\pmod{2\pi}.
		\label{eq:appC_gamma_beta}
	\end{align}
	The two inversion-invariant points of the compactified loop are
	\begin{equation}
		\theta=0,\pi.
	\end{equation}
	At $\theta=\pi$, the sewing relation directly gives a parity eigenvalue equation. At the identified point $\theta=0\equiv2\pi$, the same follows by taking the endpoint limit and using the periodic gauge:
	\begin{equation}
		P\ket{u(\theta_*)}
		=
		e^{i\beta(\theta_*)}
		\ket{u(\theta_*)}
		=
		\xi(\theta_*)\ket{u(\theta_*)},
		\qquad
		\theta_*=0,\pi,
		\label{eq:appC_parity_eigen_theta}
	\end{equation}
	where
	\begin{equation}
		\xi(\theta_*)=\pm1.
	\end{equation}
	Therefore,
	\begin{equation}
		e^{i\beta(0)}=\xi(0),
		\qquad
		e^{i\beta(\pi)}=\xi(\pi).
		\label{eq:appC_beta_xi}
	\end{equation}
	Exponentiating Eq.~\eqref{eq:appC_gamma_beta}, one obtains
	\begin{equation}
		e^{i\gamma}
		=
		e^{i\beta(0)}e^{-i\beta(\pi)}
		=
		\xi(0)\xi(\pi)^{-1}.
		\label{eq:appC_gamma_xi_intermediate}
	\end{equation}
	Since $\xi(\pi)=\pm1$, one has $\xi(\pi)^{-1}=\xi(\pi)$, and hence
	\begin{equation}
		e^{i\gamma}
		=
		\xi(0)\xi(\pi).
		\label{eq:appC_gamma_xi}
	\end{equation}
	For the branch considered in the main text, the states at the two inversion-invariant points satisfy
	\begin{equation}
		\ket{u(0)}= \ket{s},
		\qquad
		\ket{u(\pi)}\propto\ket{D_-}.
		\label{eq:appC_endpoint_states}
	\end{equation}
	At $\Delta=0$, the other two eigenvalues are $E_\pm=(U\pm\sqrt{U^2+16t^2})/2$. Since $E_-<U<E_+$ for $t>0$, $\ket{D_-}$ is the middle eigenstate for every finite $U$. Using
	\begin{equation}
		P\ket{s}=+\ket{s},
		\qquad
		P\ket{D_-}=-\ket{D_-},
	\end{equation}
	we obtain
	\begin{equation}
		\xi(0)=+1,
		\qquad
		\xi(\pi)=-1.
		\label{eq:appC_endpoint_parities}
	\end{equation}
	Thus
	\begin{equation}
		e^{i\gamma}
		=
		-1,
		\qquad
		\gamma
		=
		\pi
		\pmod{2\pi}.
		\label{eq:appC_gamma_pi}
	\end{equation}
	This proves that the geometric phase of the tracked compactified branch is fixed to
	$\pi$ by inversion symmetry.

    \section{Real-structure protection and the first Stiefel--Whitney invariant}
	\label{app:real_structure}

	We now give a complementary explanation of why the geometric phase of the
	adiabatic exchange branch is topologically protected by the real structure of
	the Hamiltonian. We use the singlet Hamiltonian and compactification defined in Eqs.~\eqref{eq:appC_H_delta} and~\eqref{eq:appC_compactification}, respectively.
    The followed branch and its closed eigenprojector loop are those established in Appendix~\ref{app:geometric_phase_inversion}.
	Let $K$ denote complex conjugation in this fixed basis,
	\begin{equation}
		K(c_+,c_-,c_s)^{\mathsf T}=(c_+^*,c_-^*,c_s^*)^{\mathsf T}.
	\end{equation}
	Here $c_+$, $c_-$, and $c_s$ are the complex expansion coefficients in
	the singlet basis defined in the main text, ${\mathsf T}$ denotes
	transpose, and $*$ denotes complex conjugation. Since every matrix
	element of $H_{\mathcal S}(\theta)$ is real,
	\begin{equation}
		K H_{\mathcal S}(\theta)K^{-1}=H_{\mathcal S}(\theta),
		\qquad K^2=1.
		\label{eq:real_structure}
	\end{equation}
	The reality condition allows a normalized real eigenvector to be chosen
	continuously on any sufficiently small parameter interval. It does not,
	however, guarantee that one real eigenvector can be chosen continuously
	with the same sign around the entire sweep; this is the global sign
	question addressed below. We therefore cover the sweep
	by overlapping intervals, labeled by $a$ and $b$, and choose normalized
	real eigenvectors $\ket{u_R^{(a)}}$ and $\ket{u_R^{(b)}}$ on them, where
	$R$ denotes a real gauge. On an overlap, both vectors span the same
	one-dimensional nondegenerate eigenspace. They must therefore be
	proportional, and reality together with normalization restricts the
	proportionality factor to a sign $g_{ab}=\pm1$:
	\begin{equation}
		\ket{u_R^{(b)}}=g_{ab}\ket{u_R^{(a)}},
		\qquad g_{ab}=\pm1.
		\label{eq:real_transition}
	\end{equation}
    Each transition sign is constant on every connected component of a patch overlap.
	The endpoint identification in Eq.~\eqref{eq:projector_limits} defines the closed projector loop over $S^1$. Its real eigenspaces,
    \begin{equation}
    \mathcal L_\theta
    =
    \{v\in\mathbb R^3:
    P_{\rm ad}(\theta)v=v\},
    \label{eq:real_bundle}
    \end{equation}
    form a real line bundle $\mathcal L$ over this circle.

	For a real line bundle over a closed one-dimensional base, the relevant
	topological invariant is the first Stiefel--Whitney class
	$w_1(\mathcal L)$~\cite{Ahn2019}. The structure group is
	$O(1)=\{\pm1\}$, representing the two possible sign changes of a real
	line. The class $w_1(\mathcal L)$ is the $\mathbb Z_2$-valued obstruction
	to choosing a continuous orientation around the loop: it is trivial when
	a transported real eigenvector returns with the same sign and nontrivial
	when it returns with the opposite sign. Its value on the complete loop is
	the first Stiefel--Whitney number
	\begin{equation}
		\nu_1=\left\langle w_1(\mathcal L),[S^1]\right\rangle
		\in\mathbb Z_2,
		\qquad
		(-1)^{\nu_1}=\prod_{\rm loop}g_{ab}.
		\label{eq:sw_number}
	\end{equation}
	Here $[S^1]$ denotes one complete traversal of the closed parameter loop,
	and the brackets mean evaluating $w_1(\mathcal L)$ on that traversal.
	The product follows successive patch changes during one traversal, including the closing transition. It is invariant under independent sign choices for the local real eigenvectors.

	For a normalized real eigenvector, $\partial_\theta\braket{u_R}{u_R}=2\braket{u_R}{\partial_\theta u_R}=0$.
    Hence the Berry connection vanishes within each real gauge patch:
	\begin{equation}
		\mathcal A_R(\theta)
		=i\braket{u_R(\theta)}{\partial_\theta u_R(\theta)}=0.
		\label{eq:real_berry_connection}
	\end{equation}
	Parallel transport within each patch therefore contributes no phase, and the Berry holonomy is the product of the transition signs around the loop:
	\begin{equation}
		e^{i\gamma}
		=\prod_{\rm loop}g_{ab}
		=(-1)^{\nu_1},
		\qquad
		\gamma=\pi\nu_1\pmod{2\pi}.
		\label{eq:sw_berry_relation}
	\end{equation}
	A nonzero $\nu_1$ obstructs a continuous periodic real eigenvector. A complex periodic gauge remains available, as used in Appendix~\ref{app:geometric_phase_inversion}.

	To determine $\nu_1$, first consider $U=0$. A normalized real eigenvector of the middle branch is
    \begin{equation}
    \ket{u_R^{(0)}(\theta)}
    =
    \cos\frac{\theta}{2}\ket{s}
    -
    \sin\frac{\theta}{2}\ket{D_-}.
    \label{eq:real_zeroU_state}
    \end{equation}
    It is continuous on the cut interval $[0,2\pi]$ and evolves from $\ket{s}$ to $-\ket{s}$. Now vary the interaction continuously as $U_\lambda=\lambda U$, with $0\leq\lambda\leq1$, at fixed $t>0$. The middle branch remains isolated, the Hamiltonian remains real, and the endpoint projector in Eq.~\eqref{eq:projector_limits} is unchanged. The discrete endpoint sign therefore cannot change during this deformation. Thus, for every finite $U$,
	\begin{equation}
		\ket{u_R(2\pi)}=-\ket{u_R(0)}.
	\end{equation}
	Equivalently, the product of its real transition signs is $-1$.
	Consequently, this branch has $\nu_1=1$ and $\gamma=\pi\pmod{2\pi}$.

    \section{Numerical extraction of the geometric phase}
    \label{app:geometric_phase_numerics}

    \begin{figure}[t]
    \centering
    \includegraphics[width=\linewidth]{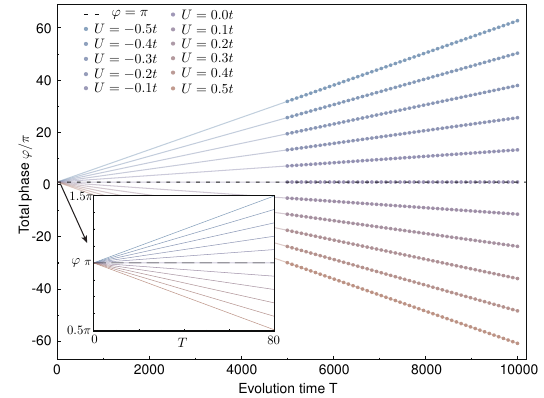}
    \caption{Numerical extraction of the geometric phase for a constant-tunnelling bias sweep with $\Delta_{\max}=40t$. The main panel shows the unwrapped singlet-return phase $\varphi_{\mathrm{tot}}/\pi$ versus the sweep duration $T$, measured in units of $\hbar/t$, for $U/t=-0.5,-0.4,\ldots,0.5$. Markers display a subset of the simulated data. Solid lines are linear fits to all sampled durations in $Tt/\hbar\in[5000,10000]$, extended toward $T=0$. The inset magnifies the same extrapolated lines near their intercepts at $\varphi_{\mathrm{tot}}\simeq\pi$.}
    \label{fig:geometric_phase_numerics}
    \end{figure}

    To test the phase quantization at finite interactions, we evolve the singlet-sector Hamiltonian in Eq.~\eqref{eq:appC_H_delta} and separate the geometric and dynamical contributions to the accumulated phase. The tunnelling amplitude $t$ is held constant, and the bias follows the linear sweep
    \begin{equation}
    \Delta(\tau)=\Delta_{\max}\left(\frac{2\tau}{T}-1\right),
    \qquad 0\leq\tau\leq T,
    \label{eq:phase_numerics_sweep}
    \end{equation}
    with $\Delta_{\max}=40t$. Energies and times are reported in units of $t$ and $\hbar/t$, respectively. The initial state is $\ket{\psi(0)}=\ket{s}$, which approximates the followed middle-branch eigenstate at the initial large negative bias.
    For each interaction strength $U/t=-0.5,-0.4,\ldots,0.5$, we solve the time-dependent Schrödinger equation for sweep durations $Tt/\hbar\in[5000,10000]$. The evolution uses a norm-preserving Cayley propagator with $N=200000$ uniform time steps. After each sweep, we extract the phase of the singlet-return amplitude,
    \begin{equation}
    \varphi_{\mathrm{tot}}(T,U)
    =\arg\!\left[\braket{s}{\psi(T,U)}\right].
    \label{eq:phase_numerics_return}
    \end{equation}

    Changing $T$ rescales the traversal rate while preserving the parameter path. In the adiabatic limit, the dynamical phase therefore depends linearly on $T$:
    \begin{equation}
    \varphi_{\mathrm{dyn}}(T,U)
    =-\frac{T}{\hbar}\int_0^1
    E_{\mathrm{mid}}\!\left(U,\Delta_{\max}(2x-1)\right)\,dx,
    \label{eq:phase_numerics_dynamic}
    \end{equation}
    where $E_{\mathrm{mid}}(U,\Delta)$ is the instantaneous eigenenergy of the followed middle branch. For each fixed $U$, we unwrap the simulated phases in increasing order of $T$ and fit them to
    \begin{equation}
    \varphi_{\mathrm{tot}}(T,U)
    \simeq \omega_{\mathrm{dyn}}(U)T+\gamma_{\mathrm{fit}}(U).
    \label{eq:phase_numerics_fit}
    \end{equation}
    Both the slope and the intercept are free fit parameters. The slope describes the dynamical contribution, while $\gamma_{\mathrm{fit}}$ estimates the geometric phase in the adiabatic, large-bias regime. The fitted intercept is reported modulo $2\pi$ on the branch closest to $\pi$.

    Figure~\ref{fig:geometric_phase_numerics} shows linear phase accumulation, with opposite dynamical slopes for positive and negative interactions and a constant phase for $U=0$. The fitted intercepts agree with $\pi$ within numerical accuracy throughout the sampled interaction range, consistent with the quantization derived in Appendices~\ref{app:geometric_phase_inversion} and~\ref{app:real_structure}. These intercepts are obtained by extrapolating the slow-sweep fits toward $T=0$.

    \section{Lattice model mapping}
	\label{app:lattice_mapping}
	
	In this Appendix, we show how the three-level Hamiltonian Eq.~\eqref{eq:singlet_bloch_hamiltonian} used in the
	main text can be formally interpreted as a Bloch Hamiltonian of an auxiliary
	one-dimensional stub lattice model with long-range hopping. This mapping provides an intuitive synthetic-lattice picture
	for the origin of the topological protection and for the role of inversion symmetry.
	
	For the purpose of constructing an auxiliary lattice model, we can formally identify $k\equiv\theta$. Then the initial Hamiltonian becomes a three-band Bloch Hamiltonian,
	\begin{equation}
		H_{\mathcal S}(k)
		=
		\begin{pmatrix}
			U & h_{AB}(k) & -2t\\
			h_{AB}(k) & U & 0\\
			-2t & 0 & 0
		\end{pmatrix},
		\label{eq:app_Hk}
	\end{equation}
	where
	\begin{equation}
		h_{AB}(k)
		=
		2\Delta(k)
		=
		-2t\cot\frac{k}{2}.
		\label{eq:app_hAB}
	\end{equation}
	We now introduce three sites in each unit cell:
	\begin{equation}
		\ket{A}\equiv \ket{D_+},
		\qquad
		\ket{B}\equiv \ket{D_-},
		\qquad
		\ket{C}\equiv \ket{s}.
		\label{eq:app_ABC_mapping}
	\end{equation}
	In this notation, the Hamiltonian in Eq.~\eqref{eq:app_Hk} describes an auxiliary
	one-dimensional stub lattice with three sublattices per unit cell. The $A$ and $B$ sites have interaction energy $U$. There is an intracell hopping between $A$ and $C$, there is no direct hopping between $B$ and $C$. The $A$--$B$ coupling is encoded in the momentum-dependent term $h_{AB}(k)$.
	
	We next derive the real-space hopping amplitudes corresponding to the
	momentum-dependent $A$--$B$ coupling. Let $\ket{\alpha,n}$ denote site
	$\alpha=A,B,C$ in unit cell $n\in\mathbb{Z}$. We use the Bloch convention
	\begin{equation}
		\ket{\alpha,k}
		=
		\sum_{n\in\mathbb{Z}}
		e^{-ikn}
		\ket{\alpha,n}.
	\end{equation}
	With this convention, the $A$--$B$ Bloch matrix element is expanded as
	\begin{equation}
		h_{AB}(k)
		=
		\sum_{r\in\mathbb{Z}}
		T_{AB}(r)e^{ikr},
		\label{eq:app_hAB_Fourier}
	\end{equation}
	where $T_{AB}(r)$ denotes hopping from a $B$ site in unit cell $n$ to an
	$A$ site in unit cell $n+r$.
	
	The Fourier coefficient is
	\begin{equation}
		T_{AB}(r)
		=
		\frac{1}{2\pi}
		\int_{0}^{2\pi}
		h_{AB}(k)e^{-ikr}\,dk .
		\label{eq:app_TAB_def}
	\end{equation}
	Using
	\begin{equation}
		\cot\frac{k}{2}
		=
		2\sum_{m=1}^{\infty}\sin(mk),
		\label{eq:app_cot_series}
	\end{equation}
	under the usual principal-value regularization of the singular point at $k=0$,
	we obtain
	\begin{align}
		h_{AB}(k)=
		-2t\cot\frac{k}{2}
		\nonumber
		=
		-4t\sum_{m=1}^{\infty}\sin(mk)
		\nonumber \\
		=
		2it\sum_{m=1}^{\infty}
		\left(
		e^{imk}-e^{-imk}
		\right).
		\label{eq:app_hAB_series}
	\end{align}
	Therefore,
	\begin{equation}
		T_{AB}(r)
		=
		\begin{cases}
			2it, & r=1,2,3,\cdots,\\[3pt]
			-2it, & r=-1,-2,-3,\cdots,\\[3pt]
			0, & r=0.
		\end{cases}
		\label{eq:app_TAB_result}
	\end{equation}                                                                         
	The coefficients satisfy
	\begin{equation}
		T_{AB}(-r)=T_{AB}(r)^*.
	\end{equation}
	These coefficients are purely imaginary and odd in $r$, which ensures hermiticity.
	
	The auxiliary real-space Hamiltonian can therefore be written as
	\begin{align}
		H_{\rm lat}
		&=
		\sum_{n}
		\left[
		U a_n^\dagger a_n
		+
		U b_n^\dagger b_n
		\right]
		-2t\sum_n
		\left(
		a_n^\dagger c_n
		+
		c_n^\dagger a_n
		\right)
		\nonumber\\
		&\quad
		+
		\sum_{n,r\neq 0}
		\left[
		T_{AB}(r)
		a_{n+r}^\dagger b_n
		+
		T_{AB}(r)^*
		b_n^\dagger a_{n+r}
		\right],
		\label{eq:app_real_space_H}
	\end{align}
	where $a_n$, $b_n$, and $c_n$ annihilate particles on the $A$, $B$, and $C$ sites
	of unit cell $n$, respectively.
	This synthetic lattice has long-range hopping in the $A$--$B$ channel, a direct consequence of the relation
	$\Delta(k)=-t\cot(k/2)$. 
	A schematic representation of this auxiliary lattice is shown in Fig.~\ref{fig:lattice_schematic}(a), illustrating the sublattices $A$, $B$, $C$ and the hoppings.
	
	The auxiliary lattice introduced above provides an intuitive picture of the parity-protected Zak phase. 
	The lattice model preserves inversion symmetry, $P H_{\mathcal S}(k) P^{-1} = H_{\mathcal S}(-k)$, consistent 
	with the original double-well Hamiltonian. At the inversion-invariant points 
	$k=0,\pi$, the eigenstate parity eigenvalues reproduce the Zak phase 
	$\gamma_{\rm Zak}=\pi$. The synthetic lattice thus provides a real-space 
	interpretation of the geometric phase accumulation. 

    \section{Numerical methods for the robustness comparison}
    \label{app:robustness}

    For all three gate mechanisms considered in Fig.~\ref{fig:gate_robustness}, we calculate the dynamics generated by the Hamiltonian in Eq.~\eqref{eq:effective_exchange_hamiltonian}. The nominal control parameters used in Fig.~\ref{fig:gate_robustness} are listed in Table~\ref{tab:gate_working_points}. These working points target population return and a singlet phase of $\pi/2$ relative to the triplets, without additional $2\pi$ windings.

    \begin{table}[t]
    \centering
    \small
    \setlength{\tabcolsep}{4pt}
    \begin{ruledtabular}
    \begin{tabular}{lccc}
        Protocol & $U/t_0$ & $\Delta_{\mathrm{end}}/t_0$
        & $Tt_0/\hbar$ \\
        \hline
        Adiabatic exchange    & 0.27     & 3.564384 & 24.171479 \\
        Nonadiabatic exchange & 2.033677 & 0        & 3.328297  \\
        Superexchange        & 10       & 0        & 13.236165 \\
    \end{tabular}
    \end{ruledtabular}
    \caption{Nominal working points for the three protocols compared in Fig.~\ref{fig:gate_robustness}. The tunnelling follows the Blackman envelope~\cite{Harris1978},
    $t(\tau)=t_0[0.42-0.5\cos(2\pi\tau/T)+0.08\cos(4\pi\tau/T)]$,
    with $0\leq\tau\leq T$.
    Here $t_0$ is the nominal peak tunnelling amplitude, $T$ is the duration of one gate, and $\Delta_{\mathrm{end}}$ is the bias amplitude.
    Each gate uses the same bias sweep given in Eq.~\eqref{eq:repeated_bias_sweeps}; $\Delta_{\mathrm{end}}=0$ gives a symmetric double well throughout the sequence.}
    \label{tab:gate_working_points}
    \end{table}

    \begin{figure}[!t]
    \centering
    \includegraphics[width=\linewidth]{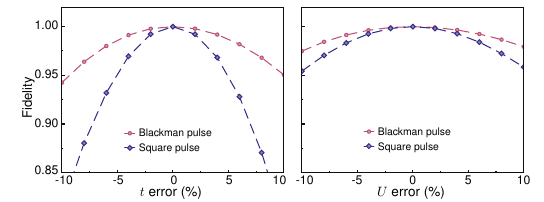}
    \caption{Static-error response of nonadiabatic exchange with square and Blackman tunnelling pulses.
    The square pulse has $t(\tau)=t_0$ for $0<\tau<T$, with
    $U/t_0=4/\sqrt{3}$ and
    $Tt_0/\hbar=\pi\sqrt{3}/4$.
    The Blackman working point is listed in Table~\ref{tab:gate_working_points}.
    Both protocols use $\Delta=0$.
    The panels show $F_8^{1/8}$ under static tunnelling errors (left) and interaction errors (right), with each pulse duration held at its nominal value.}
    \label{fig:first nonadiabatic}
    \end{figure}

    \begin{figure}[!t]
    \centering
    \includegraphics[width=\linewidth]{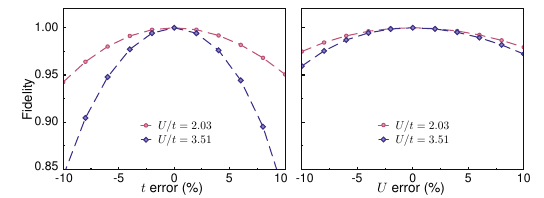}
    \caption{Static-error response at the first two nonadiabatic return points for the Blackman envelope.
    The first point is listed in Table~\ref{tab:gate_working_points}; the second uses
    $U/t_0=3.509544$ and $Tt_0/\hbar=5.179938$.
    Both use $\Delta=0$ and target the principal singlet phase of $\pi/2$.
    The panels show $F_8^{1/8}$ under static tunnelling errors (left) and interaction errors (right).}
    \label{fig:two nonadiabatic}
    \end{figure}

    For adiabatic exchange, each gate repeats the same bias sweep. Writing $\tau$ for the local time within gate $j$, we use
    \begin{equation}
    \Delta_j(\tau)
    =\Delta_{\mathrm{end}}\left(\frac{2\tau}{T}-1\right),
    \qquad j=1,\ldots,8.
    \label{eq:repeated_bias_sweeps}
    \end{equation}
    Every sweep runs from $-\Delta_{\mathrm{end}}$ to $+\Delta_{\mathrm{end}}$, and the tunnelling envelope is repeated identically. Nonadiabatic exchange and superexchange use $\Delta_j=0$ throughout.

    Static control errors are introduced as fractional offsets from the nominal parameters,
    \begin{equation}
    t_{\boldsymbol{\epsilon}}(\tau)
    =(1+\epsilon_t)t(\tau),
    \qquad
    U_{\boldsymbol{\epsilon}}
    =(1+\epsilon_U)U,
    \label{eq:static_control_errors}
    \end{equation}
    where $\boldsymbol{\epsilon}=(\epsilon_t,\epsilon_U)$.
    The tunnelling-error curves vary $\epsilon_t$ with $\epsilon_U=0$, while the interaction-error curves vary $\epsilon_U$ with $\epsilon_t=0$.
    Each offset is held constant throughout all eight gates.
    The pulse durations and programmed bias waveforms remain fixed, and the working points are not recalibrated when an error is applied.
    Energies and times remain expressed in units of the nominal $t_0$ and $\hbar/t_0$, respectively.

    Starting from the input $\ket{\Psi_0}=\ket{\uparrow,\downarrow}$ used in the main text, we propagate the state through eight consecutive identical gates. The full complex amplitudes, including any residual doublon components, are carried coherently from one gate to the next. Denoting the single-gate propagator by $\mathcal V(\boldsymbol{\epsilon})$, the complete sequence propagator is
    \begin{equation}
    \mathcal V_8(\boldsymbol{\epsilon})
    =\mathcal V(\boldsymbol{\epsilon})^8.
    \label{eq:repeated_sequence_propagator}
    \end{equation}
    Eight ideal $\sqrt{\mathrm{SWAP}}$ operations return the chosen input to itself, up to a global phase. The sequence fidelity is therefore
    \begin{equation}
    F_8(\boldsymbol{\epsilon})
    =
    \left|
    \bra{\Psi_0}
    \mathcal V_8(\boldsymbol{\epsilon})
    \ket{\Psi_0}
    \right|^2.
    \label{eq:eight_gate_fidelity}
    \end{equation}
    The overlap is evaluated without projection or renormalization onto the singly occupied subspace.
    Since the triplet component is stationary, this expression can also be written as
    \begin{equation}
    F_8
    =
    \frac{1}{4}\left|1+a_8\right|^2,
    \qquad
    a_8=
    \bra{s}\mathcal V_8(\boldsymbol{\epsilon})\ket{s}.
    \label{eq:eight_gate_singlet_amplitude}
    \end{equation}
    It therefore includes both the accumulated singlet--triplet phase error and residual doublon population at the end of the sequence.

    We note the figures report $F_8^{1/8}$ as a common measure for comparing the eight-gate sequences.
    Static errors accumulate coherently, so $F_8$ does not generally factorize into the eighth power of the single-gate fidelity.

    Figures~\ref{fig:first nonadiabatic} and~\ref{fig:two nonadiabatic} provide supplementary comparisons of pulse shapes and nonadiabatic working points. Both use the static-error prescription and sequence fidelity defined above.

	\bibliography{references}

\end{document}